\documentclass[aps,twocolumn,tightenlines,nofootinbib,superscriptaddress]{revtex4-1}
\usepackage{mathrsfs}
\usepackage{amsfonts}
\usepackage{mathtools}
\usepackage{setspace}
\usepackage{cellspace}
\usepackage{amsmath, amssymb, bm}
\usepackage[colorlinks=true,linkcolor=blue]{hyperref}
\usepackage{xcolor}
\usepackage{epsfig}
\usepackage{slashed}
\usepackage{subcaption}
\usepackage{hhline, multirow, tabularx, makecell}  %
\usepackage{dcolumn}    %
\usepackage{url}        %
\usepackage{braket}     %
\usepackage{pifont}
\usepackage{tabularx}
\renewcommand{\bra}[1]{\left<#1\left|}
\renewcommand{\ket}[1]{\right|#1\right>}

\usepackage{physics}

\begin{document}

\title{GUMP1.0---First global extraction of generalized parton distributions \\  from experiment and lattice data with NLO accuracy}
\author{Yuxun Guo}
    \email[Email: ]{yuxunguo@lbl.gov}
        \affiliation{Physics Department, University of California, Berkeley, California 94720, USA}
        \affiliation{Nuclear Science Division, Lawrence Berkeley National Laboratory, Berkeley, CA 94720, USA}
        
\author{Fatma P. Aslan}
    \email[Email: ]{fpaslan@jlab.org}
        \affiliation{Center for Nuclear Femtography, SURA, 12000 Jefferson Avenue
Newport News, VA 23606, USA}

\author{Xiangdong Ji}
    \email[Email: ]{xji@umd.edu}
        \affiliation{Maryland Center for Fundamental Physics, Department of Physics, University of Maryland,\\ 4296 Stadium Dr., College Park, MD 20742, USA}
\author{M.~Gabriel~Santiago}
    \email[Email: ]{melvin.santiago@temple.edu}
        \affiliation{Center for Nuclear Femtography, SURA, 12000 Jefferson Avenue
Newport News, VA 23606, USA}
        \affiliation{Department of Physics, Old Dominion University, Norfolk, VA 23606, USA}
        \affiliation{Jefferson Lab, Newport News, VA 23606, USA}
        \affiliation{Department of Physics, Temple University, Philadelphia, PA 19122}

\begin{abstract}
We report the first global extraction of generalized parton distributions (GPDs), GUMP1.0, by combining deeply virtual Compton scattering and $\rho$-meson production data from Jefferson Lab and Hadron-Electron Ring Accelerator with global fits of parton distribution functions, charge form factors, and lattice quantum chromodynamics simulations. Using a conformal moment space parametrization, we achieve a unified description across low- and high-$x$ regions at next to leading order (NLO) accuracy in perturbative corrections. The results provide state-of-the-art GPDs consistent with almost all known facts, enabling three-dimensional nucleon imaging in impact parameter space and, at the same time, establishing a benchmark for future theoretical and experimental studies of the nucleon structure. 
\end{abstract}

\maketitle

{\textit{Introduction---}}The multi-dimensional structures of the nucleon have drawn growing interest in the past decades as the new frontier for the study of strong interactions and quantum chromodynamics (QCD). Generalized parton distributions (GPDs) that encode the 3-dimensional (3D) distributions of partons in the nucleon have been one of the most sophisticated tools for such a purpose~\cite{Ji:1994av, Ji:1996ek, Muller:1994ses, Ji:1998pc, Goeke:2001tz, Diehl:2003ny, Belitsky:2005qn, Ji:2016djn}, together with their momentum-space counterpart---transverse momentum distributions (TMDs)~\cite{Boussarie:2023izj}. In particular, GPDs correspond to parton distributions localized in the impact parameter space~\cite{Burkardt:2000za, Burkardt:2002hr, Ji:2003ak, Belitsky:2003nz}, and contain important information about the nucleon, such as the spin and mechanical properties~\cite{Ji:1994av, Ji:1996ek, Polyakov:2002yz, Ji:2025qax}. Experimentally, GPDs can be constrained by hard exclusive scattering processes off the nucleon, such as deeply virtual Compton scattering (DVCS)~\cite{Ji:1996nm} and deeply virtual meson production (DVMP)~\cite{Radyushkin:1996ru, Collins:1996fb}, among other hard exclusive processes~\cite{Berger:2001xd, CLAS:2021lky, Belitsky:2002tf, Guidal:2002kt, Deja:2023tuc, Pedrak:2017cpp, Duplancic:2023kwe, Qiu:2022bpq, Qiu:2022pla, Qiu:2023mrm, Grocholski:2022rqj, Nabeebaccus:2023rzr, Qiu:2024mny}. Experimental studies of GPDs have been one of the crucial physics motivations for future upgrades at Jefferson Lab (JLab)~\cite{Accardi:2023chb} and the upcoming Electron-Ion Collider (EIC) in the United States~\cite{Accardi:2012qut, AbdulKhalek:2021gbh} and in China (EIcC)~\cite{Anderle:2021wcy}.

Despite their physical significance, a complete determination of even the leading-twist GPDs has remained challenging due to the well-known inverse problem~\cite{Bertone:2021yyz}. This is because measurements of DVCS and DVMP, even when combined with forward inputs, do not uniquely determine the GPDs. Meanwhile, recent novel developments have made it possible to access parton distributions directly through first-principles lattice QCD simulations~\cite {Ji:2013dva, Ji:2020ect, Alexandrou:2020zbe, Lin:2021brq}. Especially, lattice simulations mainly constrain GPDs in the regions that experiments cannot access easily~\cite{Holligan:2025baj}, and therefore provide crucial complementary input for the extraction of GPDs. In this spirit, the GPD through universal moment parametrization (GUMP) program was put forward a few years ago~\cite{Guo:2022upw, Guo:2023ahv, Guo:2024wxy}, aiming to develop advanced extractions of GPDs from global inputs.

We recognize the important and extensive efforts that have been devoted to studying exclusive processes in the GPD framework~\cite{Kumericki:2007sa, Kumericki:2009uq, Cuic:2023mki, Kumericki:2013br, Berthou:2015oaw, Dupre:2016mai, Moutarde:2019tqa, Kumericki:2016ehc, Dutrieux:2021wll, Almaeen:2024guo}, and progress has been made to extract zero-skewness GPDs from various inputs as well~\cite{Lin:2020rxa, Guo:2022upw, Goharipour:2024atx}. Nevertheless, uncovering the 3D structure of the nucleon with a unified description across all inputs requires access to the full GPDs.

In this letter, we present the first global extraction of GPDs with comprehensive inputs that cover the major experimental and lattice constraints on GPDs. Specifically, we complement the large amount of precise DVCS measurements from JLab in the relatively high-$x$ region~\cite{CLAS:2018ddh, CLAS:2021gwi, Georges:2017xjy, JeffersonLabHallA:2022pnx, CLAS:2022syx} with the Hadron-Electron Ring Accelerator (HERA) measurement of both DVCS~\cite{H1:2009wnw} and deeply virtual $\rho$-meson production (DV$\rho$P)~\cite{ZEUS:2007iet, H1:2009cml} in the low-$x$ region at next-to-leading order (NLO) accuracy of perturbative QCD corrections for the exclusive cross sections and GPD evolutions. Moreover, as GPD functions reduce to parton distribution functions (PDFs) in the forward limit and their first moments become charge form factors, we also include in our analysis the global PDFs~\cite{Cocuzza:2022jye} and nucleon charge form factors~\cite{Ye:2017gyb}. These constraints are reinforced by the most recent lattice simulations of generalized form factors and $x$-dependent GPDs at zero and non-zero skewness~\cite{Alexandrou:2019ali, Alexandrou:2021wzv, Hackett:2023rif, Bhattacharya:2023ays, Bhattacharya:2024wtg, Chu:2025kew}. With the GPD parametrization in the conformal moment space, we present a state-of-the-art extraction of GPDs across sea and valence regions. The results enable three-dimensional nucleon imaging in impact parameter space, establish a benchmark for the future study of nucleon tomography, and also provide crucial theoretical inputs for the GPD programs at JLab and the future Electron-Ion Colliders.

{\textit{Theoretical framework and GPD parametrization---}}GPDs are the generalization of PDFs with non-zero momentum transfer, defined as
\begin{equation}
    F_{q,g}\equiv \int \frac{\text{d} \lambda}{2\pi} e^{i\lambda x} \bra{P'} O_{q,g}(\lambda n) \ket{P}\ ,
\end{equation}
where the initial and final nucleon momenta are labeled as $P$ and $P'$ respectively, and the average momentum $\bar P\equiv(P+P')/2$ and momentum transfer $\Delta\equiv P'-P$ are also defined, using the notations and conventions in Ref.~\cite{Ji:1996nm}. The non-local correlators $O_{q,g}(\lambda n)$ are distributed along the direction of the light-like vector $n^2=0$ with gauge links in between. Then each GPD, collectively denoted as $F(x,\xi,t)$, depends on three variables: the average parton momentum fraction $x$, the longitudinal projection of the momentum transfer $\xi \equiv -n\cdot\Delta/(2n\cdot\bar P)$, and the momentum transfer squared $t \equiv \Delta^2 \le 0$.

Conventionally, GPDs are defined to be symmetric in $\xi$ due to parity and time-reversal symmetry. Therefore, we consider GPDs on the phase space $x\in[-1,1]$, $\xi\in[0,1]$, and $t\in(-\infty,0]$ without loss of generality. Additionally, GPDs are subject to further physical constraints, such as the endpoint constraint, kinematic constraint, and polynomiality constraint~\cite{Ji:1998pc}, among others.

It has been extensively discussed in the literature~\cite{Mueller:2005ed, Kumericki:2007sa, Kumericki:2009uq} that the conformal moments of GPDs can help one utilize these physical constraints, including the evolution properties of GPDs, to construct or parametrize GPDs. This idea naturally motivates one to consider a global analysis of GPDs by parametrizing their conformal moments. In particular, the Kumeri\v{c}ki-M\"uller (KM) model was proposed and has been successful in fitting the DVCS and DVMP measurements~\cite{Kumericki:2007sa, Kumericki:2009uq, Muller:2013jur, Mueller:2014hsa, Cuic:2023mki}, with a focus on the sea distributions. Lately, the GUMP program was put forward for the global analysis of full GPDs~\cite{Guo:2022upw, Guo:2023ahv, Guo:2024wxy}, encompassing both the sea and valence regions from extensive inputs.

In the conformal moment expansion framework, GPDs are formally expanded with respect to a set of conformal partial wave functions $p_j(x,\xi)$ as~\cite{Mueller:2005ed},
\begin{equation}
\label{eq:conformalsum}
  F(x,\xi,t) = \sum_{j=0}^{\infty} (-1)^j p_j(x,\xi) \mathcal {F}_{j}(\xi,t) \ ,
\end{equation}
where $\mathcal {F}_{j}(\xi,t)$ are the corresponding partial wave amplitudes or so-called conformal moments of GPDs. To actually construct GPDs in the $x$-space, analytical resummation of all moments $ \mathcal {F}_{j}(\xi,t)$ is required, which can usually be done through a Mellin-Barnes integral,
\begin{equation}
\label{eq:MBintegral}
F(x, \xi, t)=\frac{1}{2 i} \int_{c-i \infty}^{c+i \infty} \text{d} j \frac{p_{j}(x, \xi)}{\sin (\pi[j+1])} \mathcal{F}_{j}(\xi, t)\ .
\end{equation}
Then, GPDs can be established equivalently in the moment space. One crucial feature of constructing GPDs in moment space is that GPD moments are polynomials of $\xi$~\cite{Ji:1998pc} and can therefore be expanded in terms of $\xi$ as,
\begin{equation}
\mathcal F_j(\xi,t)=\sum_{k=0,\rm{ even}}^{j+1} \xi^{k} \mathcal{F}_{j,k}(t)\ .
\end{equation}
For the typical kinematics of interest, such as the JLab, HERA, and future EIC, $\xi\lesssim 1/2$ is small. Thus, we can truncate the above series to $k=4$ for simplicity. 

For the forward moments $\mathcal F_{j,0}(t)$ with $k=0$, we employ a simple parametrization of them as,
\begin{equation}
\label{eq:gumpform}
    \mathcal F_{j,0}(t)= \sum_{i=1}^{i_{\rm{max}}}N_{i} \frac{B\left(j+1-\alpha_{i}(t),1+\beta_{i}\right)}{B(2-\alpha_i,1+\beta_i)} R_i(t)\ , 
\end{equation}
that corresponds to the ansatzes $N_i x^{-\alpha_i}(1-x)^{\beta_i}$ in the forward limit and $x$ space, commonly used in PDF global analysis~\cite{Hou:2019efy}. The choice of $i_{\rm{max}}$ depends on the anticipated flexibility of the $x$ dependence. We set $i_{\rm{max}}=2$ for the $H$ GPD of $\bar u,\bar d,g$ to accommodate the wide coverage of $x\in[0.0005,0.6]$ and $i_{\rm{max}}=1$ for the rest. 

The $t$ dependence of the ansatzes is implemented in two ways: a non-factorized Regge trajectory: $\alpha_i(t)\equiv\alpha_i+\alpha_i' t$ with a Regge slope $\alpha'_i$, and another factorized $R(t)$ that can be parametrized separately. Recent studies of GPD parametrization with the holographic QCD model suggest a similar form~\cite{Mamo:2024vjh, Mamo:2024jwp}. Here, we additionally implement $R(t)$ to ensure sufficient flexibility in the GPD model. Specifically, we pick an exponential form $R(t)=\exp(-bt)$ for sea quarks and gluons, and a dipole form $R(t)=\left(1-t\cdot M^{-2}\right)^{-2}$ for valence quarks, to account for the different $t$ dependence of them.  

The off-forward moments $F_{j,k}(t)$ with $k\not = 0$, more specifically $k=2,4$, receive fewer constraints. Accordingly, we model them as $F_{j,k}(t)\equiv R_{k} F_{j-k,0}(t)$, proportional to the forward ones, as originally proposed in~\cite{Kumericki:2007sa, Kumericki:2009uq}. This choice can be relaxed if further constraints on GPDs are obtained from, for instance, lattice simulations. We note that the so-called $C_{q,g}$ terms are not implemented in this way, the investigation of which is left for separate future work. 

Lastly, we note that among the four leading-twist $H$, $E$, $\widetilde{H}$, and $\widetilde{E}$ GPDs, the $E$ and $\widetilde{E}$ GPDs are much less constrained due to the lack of forward constraints, e.g., from unpolarized and polarized PDFs. In particular, the flavor separation appears challenging. Therefore, we impose empirical constraints, such as $ E_g \propto H_g$, to avoid unconstrained GPDs. Again, such choices can be relaxed in the future with more constraints on the $E$ and $\widetilde{E}$ GPDs. The details of the GPD parametrization can be found in the Supplemental Material (SM) of this work, where we present all the parameters and their best-fit values, along with the full fit results.

{\textit{Setup of the analysis and results---}}In this section, we outline the setup of the global analysis and present the results. The analysis is performed using a $\chi^2$ minimization with the \texttt{iminuit} interface of \texttt{Minuit2}~\cite{iminuit, James:1975dr}. We employ the formulae in~\cite{Guo:2021gru, Guo:2022cgq} for DVCS cross sections and asymmetries with leading-twist GPDs and kinematical accuracy of twist three. For DV$\rho$P, we use leading-twist formulae, appropriate for the small-$x_B$ HERA kinematics. Both Compton (CFFs) and transition (TFFs) form factors are computed with NLO Wilson coefficients and GPD evolution, of which the expressions in conformal moment space are collected in~\cite{Kumericki:2007sa, Kumericki:2009uq, Cuic:2023mki}.

\begin{figure}[t]
    \centering
    \includegraphics[width=0.483\textwidth]{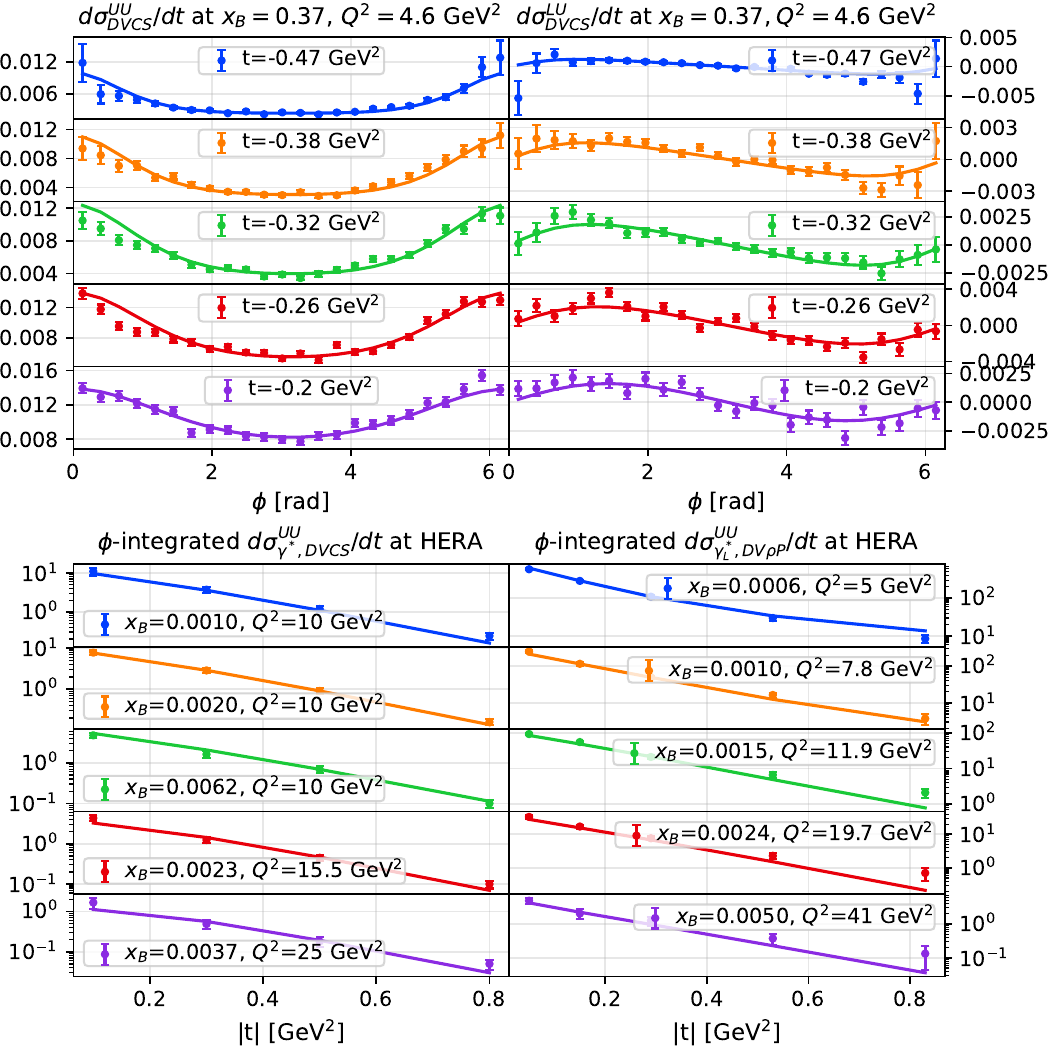}
    \caption
    {\raggedright Representative fits to differential cross sections: unpolarized DVCS at JLab (top left), beam-polarized DVCS at JLab (top right), unpolarized DVCS at HERA (bottom left), and unpolarized DV$\rho$P at HERA (bottom right). JLab data are $\phi$-differential electro-production cross sections~\cite{CLAS:2018ddh, CLAS:2021gwi, Georges:2017xjy, JeffersonLabHallA:2022pnx}, while HERA data are $\phi$-integrated photo-production cross sections~\cite{ZEUS:2007iet, H1:2009cml, H1:2009wnw}.}
    \label{fig:expfiteg}
\end{figure}

In this first analysis, some GPD-related observables have not been included for various considerations, such as the time-like Compton scattering~\cite{Berger:2001xd, CLAS:2021lky}, doubly DVCS~\cite{Belitsky:2002tf, Guidal:2002kt, Deja:2023tuc} and other 2-to-3 processes~\cite{Pedrak:2017cpp, Duplancic:2023kwe, Qiu:2022bpq, Qiu:2022pla, Qiu:2023mrm, Grocholski:2022rqj, Nabeebaccus:2023rzr, Qiu:2024mny} discussed in the literature. These could be incorporated in future work once more data are available. The exclusive $J/\psi$ productions also provide constraints on the gluon GPDs~\cite{H1:2005dtp, Ivanov:2004vd, Kowalski:2006hc, Chen:2019uit, Koempel:2011rc, Koempel:2015xol, Flett:2021ghh, Mantysaari:2021ryb, Mantysaari:2022kdm, Eskola:2022vpi, Goloskokov:2024egn, Flett:2024htj, Guo:2024wxy}. However, including these data requires extending the factorized $t$ dependence assumed in this analysis. For the same reason, we omit the lattice results on gluon gravitational form factors~\cite{Hackett:2023rif}. Furthermore, recent near-threshold $J/\psi$ measurements have advanced the study of gluon gravitational form factors~\cite{GlueX:2019mkq, Duran:2022xag, GlueX:2023pev, Guo:2021ibg, Guo:2023pqw, Guo:2023qgu, Guo:2025jiz}, but as they mainly probe the large-$\xi$ region, a dedicated analysis is again left to future work.

\begin{figure}[t]
    \centering
    \includegraphics[width=0.483\textwidth]{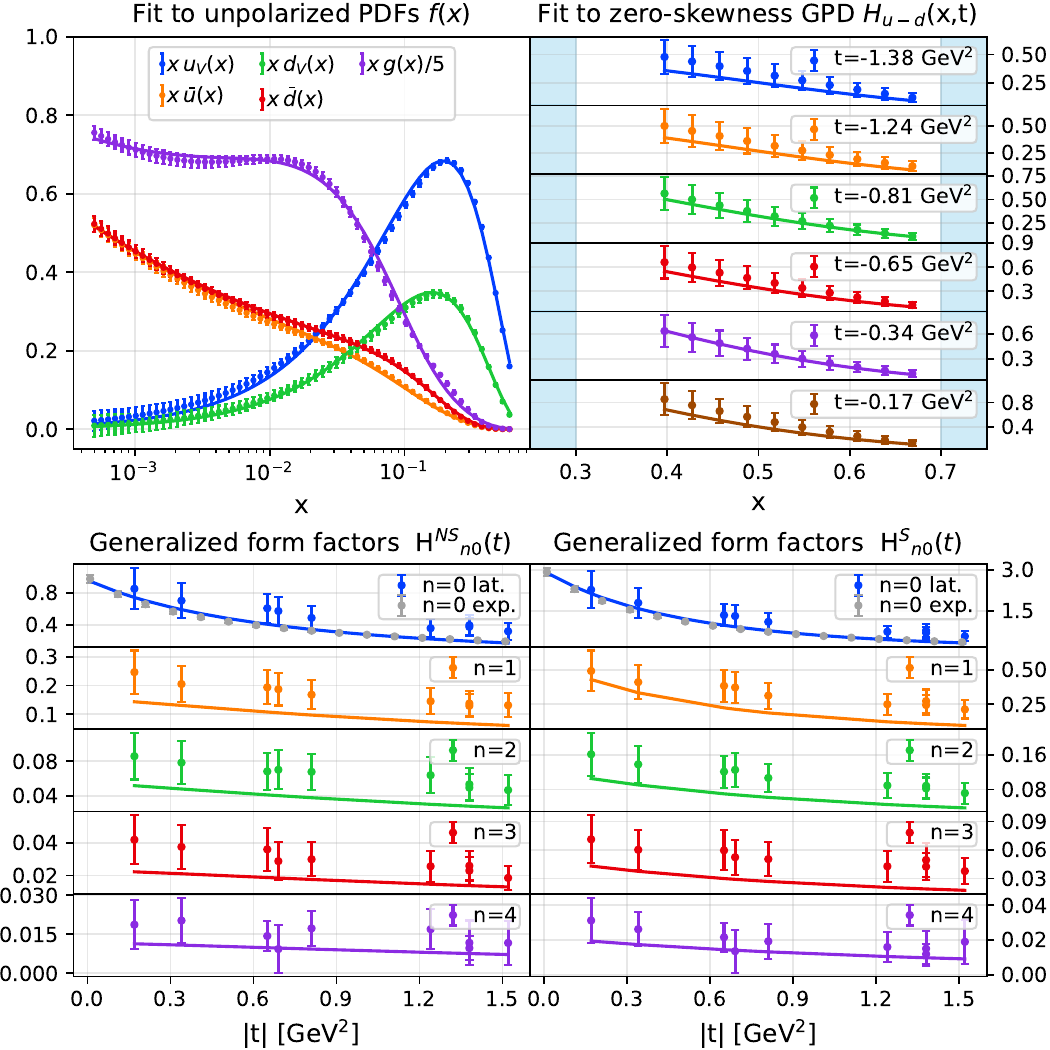}
    \caption
    {\raggedright Representative fits to global PDFs and lattice QCD simulations: fits to global unpolarized PDFs (top left)~\cite{Cocuzza:2022jye}, lattice simulations of GPDs at zero skewness (top right)~\cite{Bhattacharya:2022aob, Bhattacharya:2023jsc, Chu:2025kew}, and global charge form factor~\cite{Ye:2017gyb} and lattice simulations of generalized form factors for flavor non-singlet (NS) $H^{u-d}_{n0}(t)$ (bottom left) and flavor singlet (S) $H^{u+d}_{n0}(t)$ (bottom right)~\cite{Bhattacharya:2023ays}.
    \label{fig:latfiteg}}
\end{figure}

In Fig.~\ref{fig:expfiteg}, we present representative fits to various experimental measurements, including the unpolarized and beam-polarized DVCS differential cross sections measured at JLab~\cite{CLAS:2018ddh, CLAS:2021gwi, Georges:2017xjy, JeffersonLabHallA:2022pnx} and the $\phi$-integrated differential cross sections of DVCS and DV$\rho$P measured by H1 and ZEUS at HERA~\cite{ZEUS:2007iet, H1:2009cml, H1:2009wnw}. We also note the development of a GPD database~\cite{Burkert:2025gzu} and include the asymmetry measurements~\cite{CLAS:2022syx}. All DV$\rho$P data are converted to longitudinal cross sections using the longitudinal-to-transverse ratio $R$ extracted from experimental measurements~\cite{ZEUS:2007iet, H1:2009cml}. Generally, we observe excellent agreement with the data. In particular, we find a consistent description between the DVCS and DV$\rho$P measurements of HERA. This also aligns with the observations by the KM model~\cite{Cuic:2023mki}. The main challenge appears to be the presence of possible higher-order sinusoidal behaviors in the $\phi$ dependence of the JLab data, which could indicate higher-twist effects that are beyond the scope of this work and warrant further investigation.

In Fig.~\ref{fig:latfiteg}, we present representative fits to the globally fitted PDFs by the JLab Angular Momentum (JAM) Collaboration~\cite{Cocuzza:2022jye} as well as lattice-calculated GPDs~\cite{Bhattacharya:2022aob, Bhattacharya:2023jsc, Chu:2025kew}. We also consider the global nucleon charge form factors~\cite{Ye:2017gyb} and a systematic lattice QCD simulation of generalized form factors up to the fifth moments~\cite{Bhattacharya:2023ays}. The detailed fit results that do not fit here are collected in the SM of this letter. Such integrated constraints on GPDs across different species and flavors are essential for a robust extraction, since the experimental observables are generally insensitive to these distinctions unless measurements are performed with varying beam and target polarization configurations~\cite{Shiells:2021xqo}. This is one substantial improvement anticipated from the future EIC, which will enable high-precision measurements with different polarization configurations.

We remark that when combining the experimental data with the lattice results, a moderate relative 30\% uncertainty has been added to the reported uncertainty of the lattice results to account for systematic uncertainties (continuum limit, physical quark mass, infinite volume, excited state contamination,
large momentum extrapolation, etc.) not included in the original analysis. It serves as a rough estimate, whereas further quantitative studies of systematic uncertainties are undoubtedly crucial to analyze experimental and lattice data simultaneously, which, nevertheless, is beyond the scope of this work.

\begin{figure}[t]
    \centering
    \includegraphics[width=0.48\textwidth]{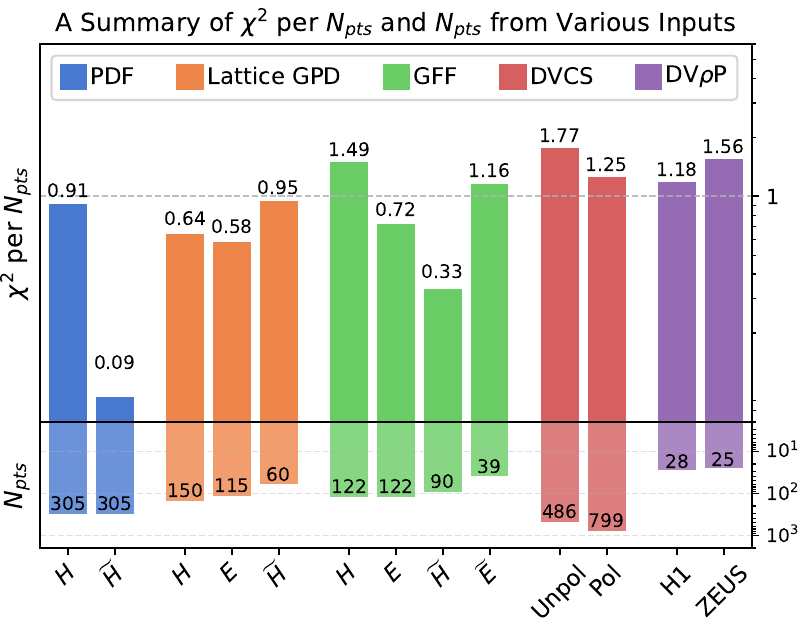}
    \caption
    {\raggedright A summary of contributions to the $\chi^2$ from all inputs used in the analysis, normalized by the number of data points $N_{\rm{pts}}$, which are shown in the lower panel.}
    \label{fig:fitsum}
\end{figure}

Finally, in Fig.~\ref{fig:fitsum}, we summarize the fit results by presenting the number of data points, $N_{\rm pts}$, from all the inputs used in this analysis, together with the corresponding $\chi^2/N_{\rm pts}$ for each of them. Overall, this analysis incorporates a total of \textbf{2,646} data points from experiments, lattice simulations, and global PDFs and form factors, yielding a $\chi^2/N_{\rm pts}$ value close to unity. The overall $\chi^2$ per degree of freedom is about \textbf{1.09}, demonstrating a good level of agreement among all the inputs.

\begin{figure}[t]
    \centering
    \includegraphics[width=0.48\textwidth]{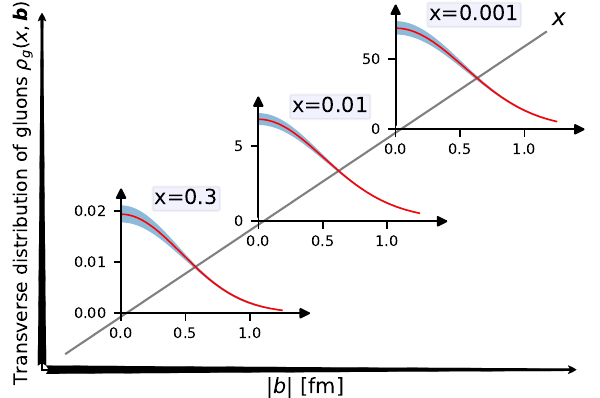}
    \caption
    {\raggedright The transverse space gluon distributions $\rho_{g}(x,\boldsymbol b)$ in an unpolarized proton with GUMP1.0 GPDs at $\mu=2$ GeV.Bands correspond to $90$\% confidence intervals considering only the Hessian errors.}
    \label{fig:gluontomo}
\end{figure}

{\textit{Extracted GPDs and proton tomography---}}One topic central to the study of GPDs is the nucleon tomography, where the impact parameter space distributions of partons in the nucleons can be written as~\cite{Burkardt:2002hr},
\begin{equation}
\begin{split}
    \rho_{q/g}(x,\boldsymbol b) = \int \frac{\text{d}^2\boldsymbol \Delta}{(2\pi)^2} e^{-i \boldsymbol{\Delta}\cdot \boldsymbol b}H_{q/g}(x,-\boldsymbol \Delta^2)  \ ,
\end{split}
\end{equation}
in terms of the corresponding zero-skewness GPDs. In this section, we present the extracted GPDs, denoted as GUMP1.0 extraction, and the proton tomography. Fig. \ref{fig:gluontomo} shows an example of the extracted intrinsic transverse space gluon distributions, $\rho_{g}(x,\boldsymbol b)$, in an unpolarized proton at $\mu=2$ GeV, which will be one of the most important goals of the future EIC~\cite{Accardi:2012qut}. We note, however, that the factorized parametrization of the $t$ dependence may limit the accuracy, as discussed above, and the errors due to parametrization bias have not been included.

Recently, the tomography of transversely polarized nucleons was revisited~\cite{Ji:2020hii, Guo:2021aik, Lorce:2021gxs, Freese:2021mzg, Panteleeva:2022khw} and the intrinsic transverse space distribution in a transversely polarized (in the $X$ direction) nucleon was defined as~\cite{Guo:2021aik},
\begin{equation}
\begin{split}
\label{rhoinXb}
    \rho_{q,\rm{In}}^{X}(x,\boldsymbol b)=&\int \frac{\text{d}^2\boldsymbol \Delta}{(2\pi)^2} e^{-i \boldsymbol{\Delta}\cdot \boldsymbol b} \Big[H_q(x,-\boldsymbol \Delta^2)\\
    &\qquad\qquad+\frac{i\Delta_y}{2M}\left(H_q+E_q\right)(x,-\boldsymbol \Delta^2)\Big]\ ,
\end{split}
\end{equation}
in terms of the zero-skewness GPDs, such that the contribution from the center-of-mass motion of the proton to its angular momentum is removed~\cite{Ji:2020hii, Guo:2021aik}. Correspondingly, the intrinsic contributions to the proton angular momentum are:
\begin{equation}
    \label{eq:amdist}
    J^{X}_{q}(x,\boldsymbol b)= \frac{\gamma}{2}( b^y \times  x P^+)\rho_{q,\rm{In}}^{X}(x,\boldsymbol{b})\ .
\end{equation}
Here $\gamma$ is the boost factor of the proton, which will be omitted hereafter to normalize the proton spin to $\hbar/2$.

\begin{figure}[t]
    \centering
    \includegraphics[width=0.46\textwidth]{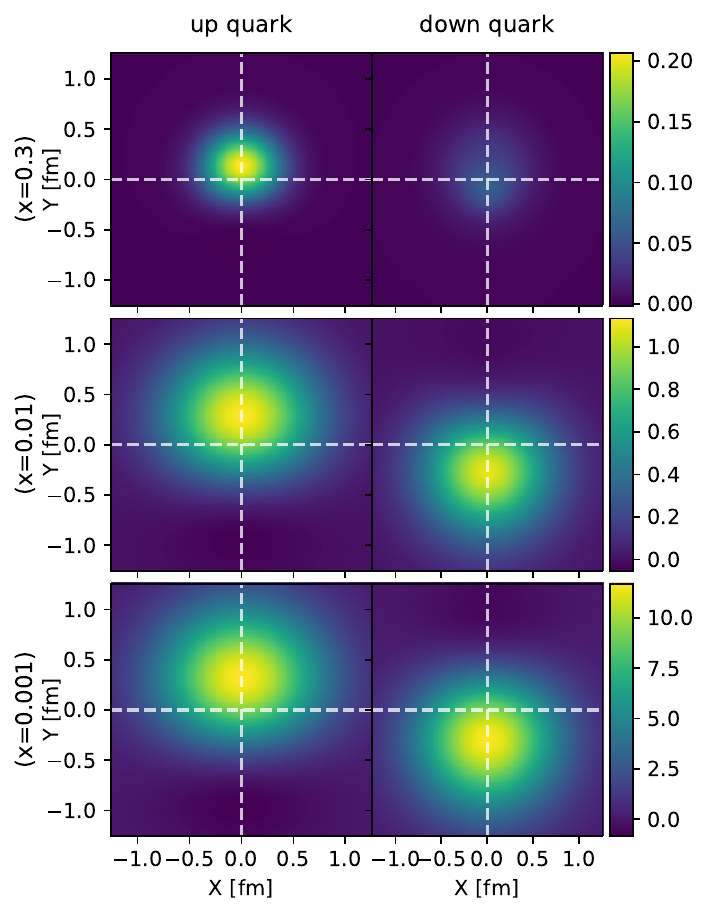}
    \caption
    {\raggedright The intrinsic transverse space distributions $\rho_{q,\rm{In}}^{X}(x,\boldsymbol b)$ of up and down quarks in a proton polarized in the $X$ direction for $x=0.3$, $0.01$, and $0.001$ with GUMP1.0 GPDs at $\mu=2$ GeV.}
    \label{fig:transplot}
\end{figure}

In Fig. \ref{fig:transplot}, we present the intrinsic transverse distributions $\rho_{q,\rm{In}}^{X}(x,\boldsymbol b)$ of up and down quarks at $x=0.3$, $0.01$, and $0.001$ respectively at the reference scale $\mu=2$ GeV. The $\rho_{q,\rm{In}}^{X}(x,\boldsymbol b)$ are calculated via a numeric Fourier transform of the extracted zero-skewness GPDs according to eq. (\ref{rhoinXb}). We label the transverse directions as $\boldsymbol b\equiv(X, Y)$ to distinguish them from the momentum fraction $x$, and assume the proton polarization along $X$.

From the intrinsic transverse quark distributions, it is clear that the quark contributions to the proton angular momentum are mainly from the up quark in the valence region. In contrast, the down quark contribution is minor and can potentially be negative. This picture gets rather different in the sea region. There appears to be a cancellation of up and down quarks' contributions to the proton angular momentum, as they are both distributed significantly off the $Y$-axis but on opposite sides. Nevertheless, as the angular momentum contributions are suppressed by $x$ in the sea region, as shown in eq.(\ref{eq:amdist}), the quark contributions to the proton angular momentum are still dominated by the up quarks. 

From the global extraction, we report the proton angular momentum decomposition to be: $J_u=0.334(9)$,  $J_d=-0.108(12)$, and $J_g=0.258(8)$ with $J_{\rm{tot}}=0.484(17)$, where uncertainties include the Hessian errors only. Comparing this to the lattice simulations~\cite{Hackett:2023rif}, which reported $J_u=0.2213(85)$, $J_d=0.0197(85)$, and $J_g=0.255(13)$, the difference is mainly caused by the global charge form factors that favor a negative $F_2^{u+d}\sim B^{u+d}_{10}(t)$. In contrast, the lattice favors positive or almost zero $B^{u+d}_{10}(t)$. The detailed comparison of the extracted $E$ form factors can be found in the SM. We leave a more careful investigation of this matter to future work.

{\textit{Conclusion---}}To summarize, we present the first global analysis of GPDs using a comprehensive set of experimental and lattice inputs within the GUMP framework and its conformal moment parametrization. This work establishes a benchmark for future studies of nucleon tomography and internal structure, while providing essential input for GPD programs at current and upcoming facilities such as JLab, EIC, and EIcC.

To demonstrate the power of the global extraction, we present the transverse spatial distributions of quarks in the proton, providing a direct tomographic image of its internal structure. This showcase highlights the unique capability of globally constrained GPDs to reveal the spatial dynamics of quarks with unprecedented fidelity.

The developments of this work will be released as open-source~\cite{Guo:2025gumpgit}, aiming to provide a long-lasting resource designed to accelerate and support all future GPD studies.

{\textit{Acknowledgments---}}We thank S. Bhattacharya, K. Cichy, M.-H. Chu, M. Constantinou, X. Gao, J. Miller, and D. Pefkou for providing the data files of their lattice calculations and correspondence. We also thank A. Camsonne and I. M. Higuera-Angulo for the useful discussions on the GPD database. We thank K. Kumericki and F. Yuan for valuable discussions. This research benefits from the open-source Gepard package~\cite{gepard}. This research is supported by the U.S. Department of Energy, Office of Science, Office of Nuclear Physics, under contract/grant/award numbers DE-AC02-05CH11231, DE-SC0020682, DE-AC05-06OR23177, and DE-FG02-97ER41028, and the Center for Nuclear Femtography, Southeastern Universities Research Association, Washington, D.C. The authors also acknowledge partial support from the U.S. Department of Energy, Office of Science, Office of Nuclear Physics, under the umbrella of the Quark-Gluon Tomography (QGT) Topical Collaboration with Award DE-SC0023646.

\bibliography{refs.bib}

\clearpage
\onecolumngrid
\setcounter{page}{1} 
\setcounter{equation}{0}
\setcounter{figure}{0}
\setcounter{table}{0}
\renewcommand{\theequation}{\rm{S.}\the\numexpr\value{equation}\relax}
\renewcommand{\thefigure}{\rm{S.}\arabic{figure}}
\renewcommand{\thetable}{\rm{S.}\arabic{table}}

\begin{center}
    {\large\textbf{Supplemental material of ``GUMP1.0---First global extraction of generalized parton distributions  from experiment and lattice data with NLO accuracy''}}\vspace{0.5cm}

    Yuxun Guo, Fatma P. Aslan, Xiangdong Ji and M.~Gabriel~Santiago
\end{center}
\vspace{0.5cm}

In this supplemental material, we provide details of the parametrization of the generalized parton distributions (GPDs) and the inputs used in our analysis. We also present comparisons between the fitted values and the inputs for most of the inputs that could not all be included in the main text, along with the corresponding best-fit parameters.

\section{Summary of GPD parametrization and inputs}

In Table. \ref{tab:gpdparam}, we collect how each GPD has been parametrizied in this work. For fully-parametrized GPDs, their forward moments $F_{j0}(t)$ are expressed as,
\begin{equation}
\label{eq:gumpform}
    \mathcal F_{j,0}(t)= \sum_{i=1}^{i_{\rm{max}}}N_{i} \frac{B\left(j+1-\alpha_{i}(t),1+\beta_{i}\right)}{B(2-\alpha_i,1+\beta_i)} R_i(t)\ , 
\end{equation}
as shown in the main text, whereas the off-forward moments $F_{j,k}(t)$ are expressed through ratio $F_{j,k}(t)\equiv R_{k} F_{j-k,0}(t)$, due to the lack of off-forward constraints on the $x$ dependence. For these GPDs that are not fully-parametrized, their forward moments are given by other GPDs they link to, e.g. , $E^{\bar u}_{j0} = R^E_{\bar u} H^{\bar u}_{j0}$. These empirical constraints are chosen since the sea regions of $E$ and $\widetilde{E}$ are typically not constrained by most existing experimental data or lattice simulations. For GPDs that are not fully-parametrized, their off-forward moments $F_{j,k}(t)$ are given in terms of ratio as well for the same reason.

\begin{table}[ht]
    \centering
    \caption{\raggedright \label{tab:gpdparam} A summary of how GPDs of different species and flavors are parametrized. Fully parametrized GPDs are expressed in terms of parametrization presented in the main text, whereas the others are linked to fully parametrized GPDs with proportional constants, following our previous setup~\cite{Guo:2023ahv}.}
    \renewcommand{\arraystretch}{2}
    \begin{tabular}
    {>{\centering\arraybackslash} p{0.3\textwidth} >{\centering\arraybackslash} p{0.18\textwidth}  >{\centering\arraybackslash} p{0.25\textwidth} >{\centering\arraybackslash} p{0.2\textwidth}} \hline 
    \hline
       GPDs species and flavors & \makecell{Fully \\parametrized}& GPDs linked to &  \makecell{Proportional \\ constants}  \\ \hline
       \makecell{$H_{u_V}$, $H_{\bar u}$, $H_{d_V}$, $H_{\bar d}$, $H_{g}$,\\ $E_{u_V}$ and $E_{d_V}$} & \ding{52} & - & - \\ \hline
       $E_{\bar u}$, $E_{\bar d}$, and $E_{g}$ & \ding{56} & $H_{\bar u}$, $H_{\bar d}$, and $H_{g}$ & $R^E_{\bar u}$, $R^E_{\bar d}$, and $R^E_{g}$ \\ \hline
        \makecell{$\widetilde{H}_{u_V}$, $\widetilde{H}_{\bar u}$, $\widetilde{H}_{d_V}$, $\widetilde{H}_{\bar d}$,\\ $\widetilde{H}_{g}$, and $\widetilde{E}_{u_V}$} & \ding{52} & - & - \\ \hline
       $\widetilde{E}_{d_V}$, $\widetilde{E}_{\bar u}$, $\widetilde{E}_{\bar d}$, and $\widetilde{E}_{g}$ & \ding{56} & $\widetilde{E}_{d_V}$, $\widetilde{E}_{\bar u}$, $\widetilde{E}_{\bar d}$, and $\widetilde{E}_{g}$ & $R^{\widetilde{E}}_{d_V}$, $R^{\widetilde{E}}_{\rm{sea}}$ \\ \hline
    \end{tabular}
\end{table}

In Table~\ref{tab:semiforwardinputs}, we provide a summary of the inputs employed in this analysis. For all lattice inputs, we introduce an additional relative uncertainty of $30\%$ on top of the reported errors to account for systematic effects not included in the original studies. While this prescription represents only a rough estimate, it appears sufficient to reconcile the discrepancies observed between lattice-calculated charge form factors and those obtained from global analyses. A comprehensive and quantitative assessment of systematic uncertainties in lattice simulations remains an important task for the future, but is beyond the scope of the present work.

Furthermore, we exclude the deeply virtual $J/\psi$ production data~\cite{H1:2005dtp} and the lattice simulations of gluon gravitational form factors~\cite{Hackett:2023rif} from our final analysis due to tensions in their $t$ dependence compared with that of deeply virtual Compton scattering and $\rho$-meson production. In principle, such differences in $t$ dependence could be accommodated within GPDs across different kinematic regions, reflecting the distinct production mechanisms among the experimental observables and the different kinematics of interest for lattice simulations. However, within the parametrization framework adopted here with a factorized residual $t$ dependence, it remains challenging to reconcile these discrepancies consistently with the rest of the dataset. A more detailed investigation of this issue is therefore deferred to future work.

\begin{table}[ht] 
    \centering
    \caption{\label{tab:semiforwardinputs} \raggedright Summary of forward, and off-forward inputs used in this analysis, including global fitted PDFs, form factors, cross sections of exclusive production processes, and lattice-calculated generalized form factors and $t$-dependent PDFs.}
    \renewcommand{\arraystretch}{2} %
    \begin{tabular}
    {>{\centering\arraybackslash} p{0.22\textwidth} >{\centering\arraybackslash} p{0.35\textwidth}  >{\centering\arraybackslash} p{0.40\textwidth}} \hline \hline
    \makecell{Category} & \makecell{Inputs} & \makecell{Comment} \\ \hline
    \makecell{(Polarized) PDFs} &  \makecell{Unpolarized and polarized \\ JAM22 PDFs~\cite{Cocuzza:2022jye} } & \makecell{For $x\in[0.0005,0.6]$, sample 60 points  \\for each flavor with positivity\\ constraints on helicity PDFs.}\\\hline
    Charge form factor & Global charge form factor~\cite{Ye:2017gyb} & \makecell{Both proton and neutron form \\  factors assuming isospin symmetry}\\ \hline
    \makecell{Generalized \\form factors} & \makecell{Isovector and isoscalar quark  \\ $A^{u\pm d}_{n}(t)$ and $B^{u\pm d}_{n}(t)$~\cite{Bhattacharya:2023ays}}  & \makecell{Up to the fifth moment} \\ \hline
    \makecell{Generalized \\ form factors~\footnote{Eventually not included in the final analysis, see the text above for explanation.}} & \makecell{Vector form factors $A^{g}_{2,0}(t)$~\cite{Hackett:2023rif}}  &  \makecell{Only leading moments\\ only gluonic ones used.} \\ \hline
    \makecell{Generalized \\form factors} & \makecell{Isovector and isoscalar quark\\axial-vector $\tilde{A}^{u\pm d}_{n}(t)$~\cite{Bhattacharya:2024wtg}}  &  \makecell{Up to the fifth moment} \\ \hline
    \makecell{Generalized \\ form factors} & \makecell{Axial-vector GFFs $\widetilde{B}^{q}_{2,0}(t)$~\cite{Alexandrou:2019ali}}  &   \makecell{Non-singlet data in the appendix used.}  \\ \hline
    \makecell{Generalized \\form factors} & \makecell{Axial-vector GFFs $\widetilde{B}^{q}_{1,0}(t)$~\cite{Alexandrou:2021wzv}}  &  \makecell{Mean values generated with the  \\
    fitted form for $u$ and $d$ quark.} \\ \hline
    \makecell{GPDs at\\ zero skewness} &  \makecell{Isovector quark GPD $H_{u-d}(x,t)$,\\ $E_{u-d}(x,t)$, and $\widetilde{H}_{u-d}(x,t)$~\cite{Bhattacharya:2022aob,Bhattacharya:2023jsc}} & \makecell{10 points in $x\in[0.3,0.7]$ are \\ sampled to avoid endpoints.} \\ \hline
    \makecell{GPDs at non-\\ zero skewness} &  \makecell{Isovector quark GPD \\ $H_{u-d}(x,t)$ and $E_{u-d}(x,t)$~\cite{Chu:2025kew}} & \makecell{10 points in $x\in[0.3,0.7]$ are \\ sampled to avoid endpoints.\\ Also exclude $|x-\xi|\le0.2$ region} \\ \hline
    \makecell{Exclusive Production \\ cross sections} & \makecell{Differential DVCS cross \\ sections from JLab~\cite{CLAS:2018ddh,CLAS:2021gwi,Georges:2017xjy,JeffersonLabHallA:2022pnx}} & \makecell{Unpolarized and polarized DVCS \\ cross sections at JLab \\ differential  in both $t$ and $\phi$}\\\hline
    \makecell{Exclusive Production \\ asymmetries} & \makecell{DVCS beam spin asymmetries\\ from JLab, mostly in~\cite{CLAS:2022syx}}& \makecell{DVCS beam spin asymmetries \\ clustered in terms of ($x_B$, $Q$ and $t$) \\ differential  in both $t$ and $\phi$}\\\hline
    \makecell{Exclusive Production \\ cross sections} & \makecell{Differential DVCS cross \\ sections from HERA~\cite{H1:2009wnw}} & \makecell{ $\phi$-integrated and $t$-differential \\ unpolarized DVCS cross sections} \\ \hline
    \makecell{Exclusive Production \\ cross sections}&\makecell{Differential DVMP cross \\ sections from HERA~\cite{ZEUS:2007iet,H1:2009cml} }  & \makecell{$\phi$-integrated and $t$-differential \\ unpolarized DV$\rho$P cross sections} \\ \hline
  \end{tabular}
\end{table}

\section{Summary of fit results and best-fit parameters}

\begin{figure}[ht]
    \centering
    \includegraphics[width=0.75\textwidth]{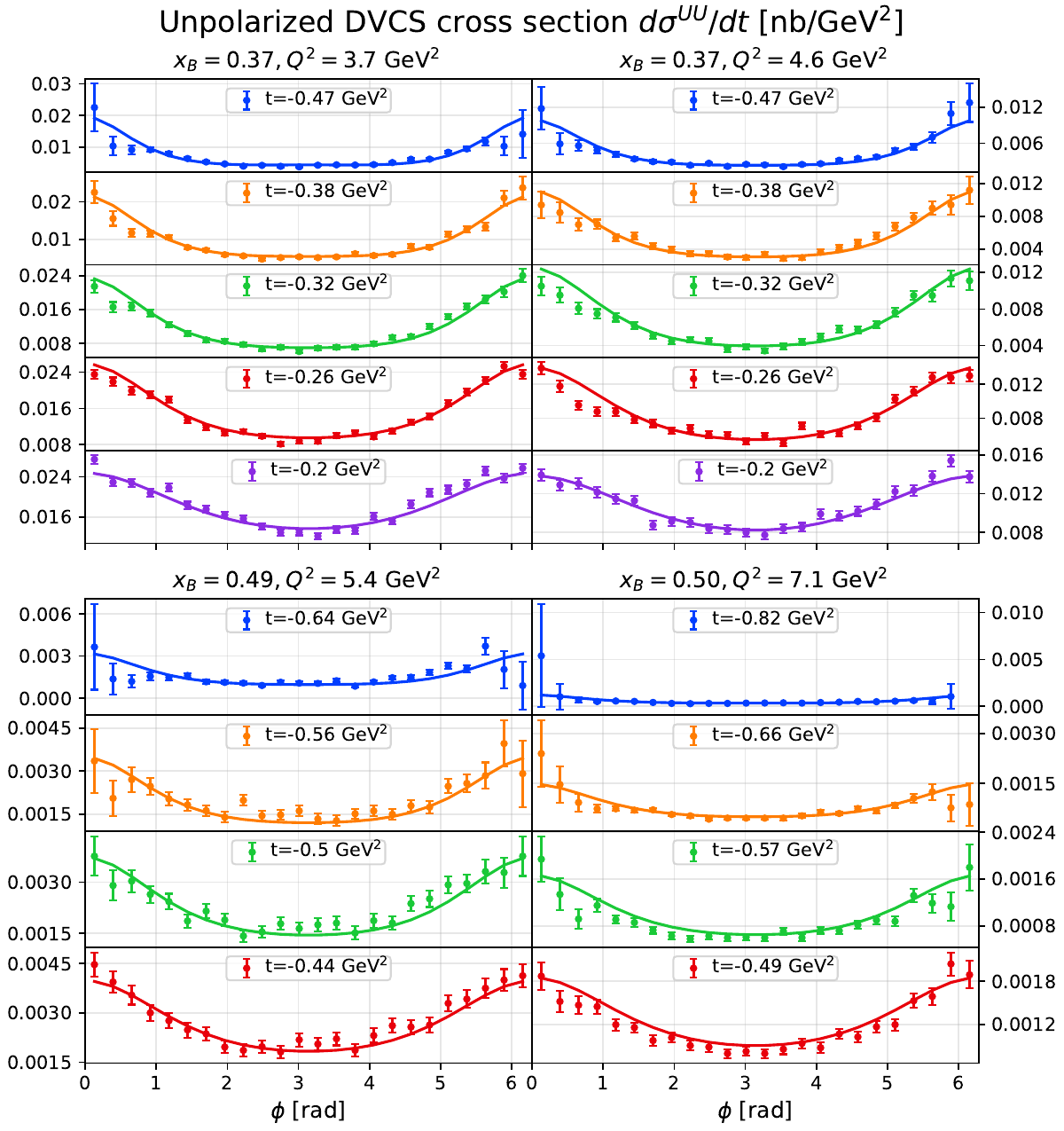}
    \caption{\raggedright Fits to the unpolarized DVCS cross sections measured at JLab.}
    \label{fig:DVCSUU}
\end{figure}

\begin{figure}[ht]
    \centering
    \includegraphics[width=0.75\textwidth]{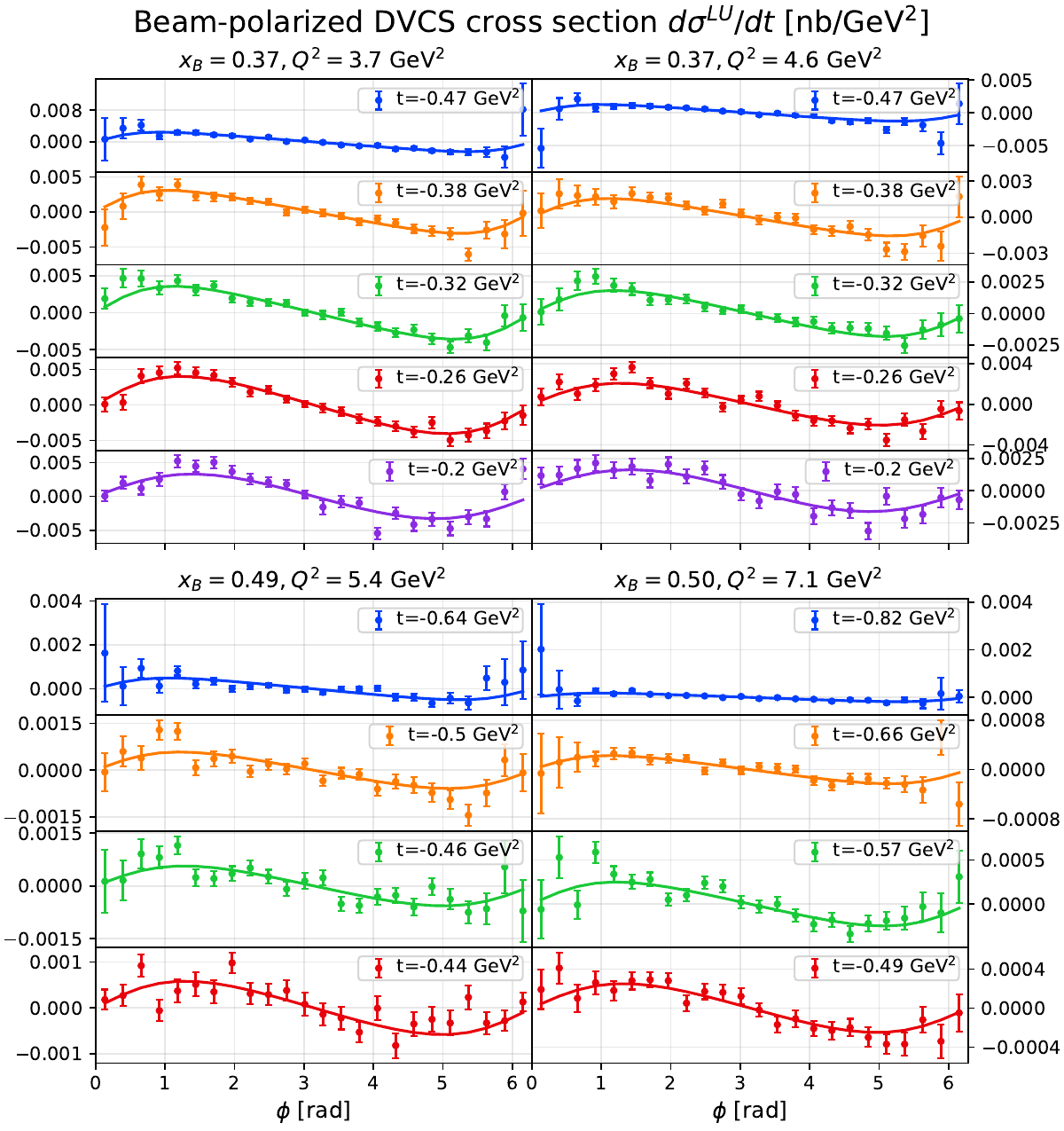}
    \caption{\raggedright Fits to the beam-polarized DVCS cross sections measured at JLab.}
    \label{fig:DVCSLU}
\end{figure}

\begin{figure}[ht]
    \centering
    \includegraphics[width=0.75\textwidth]{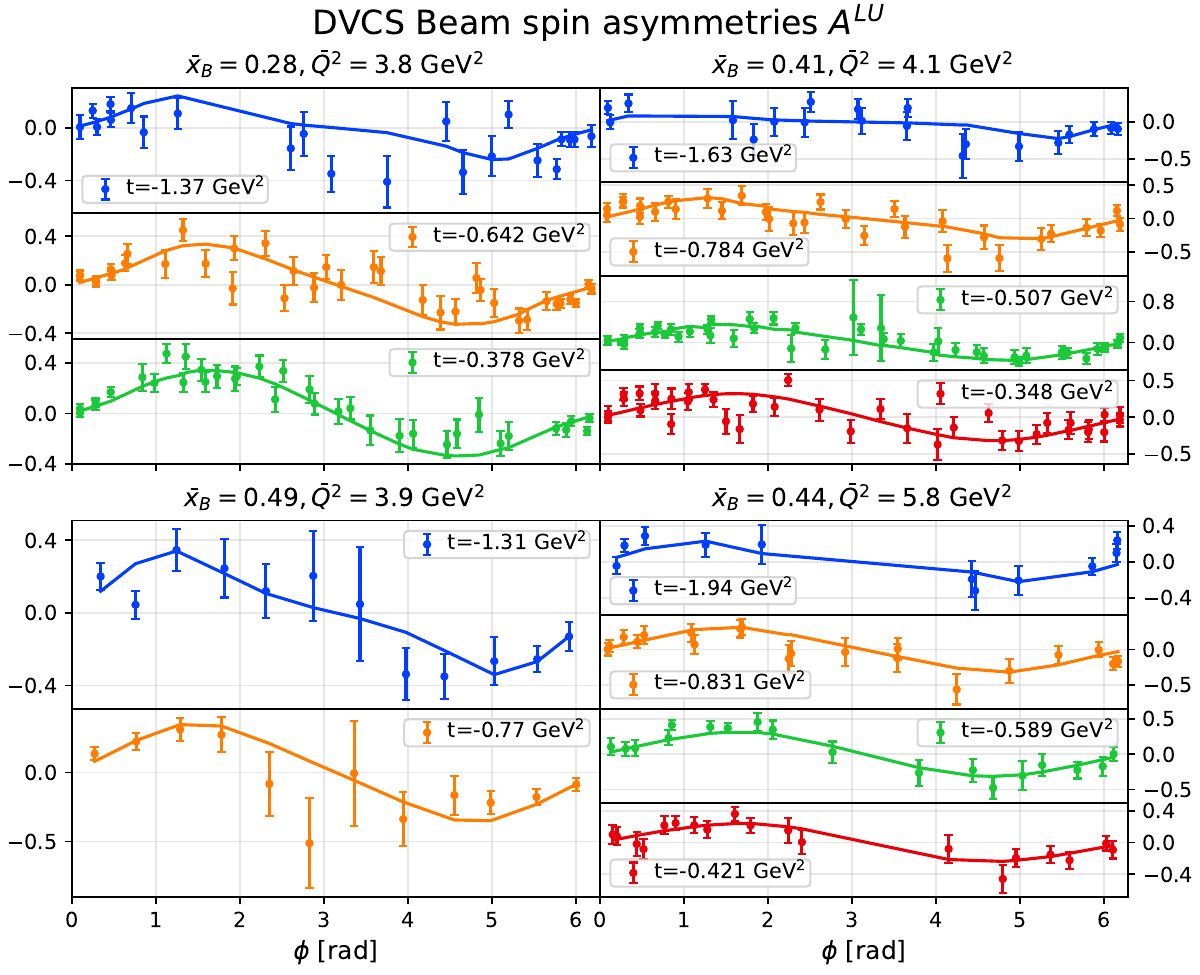}
    \caption{\raggedright Fits to the DVCS beam-spin asymmetries measured at JLab.}
    \label{fig:DVCSALU}
\end{figure}

\begin{figure}[ht]
    \centering
    \includegraphics[width=0.95\textwidth]{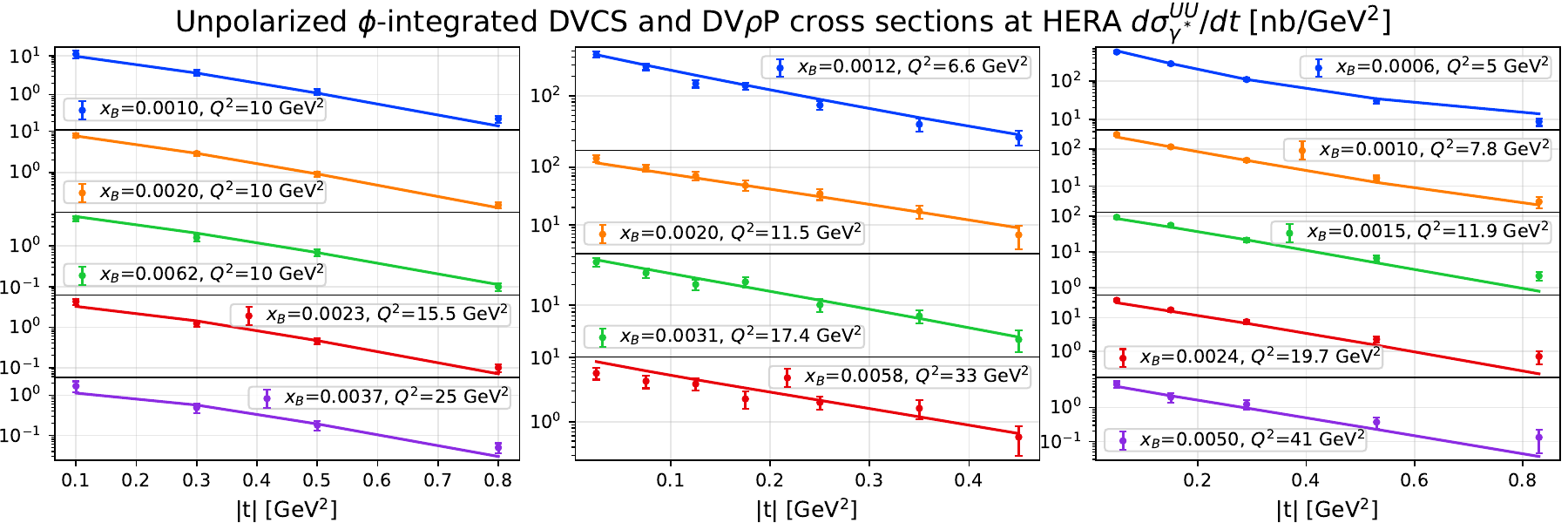}
    \caption{\raggedright Fits to the DVCS and (longitudinal) DV$\rho$P cross sections measured at HERA.}
    \label{fig:HERA}
\end{figure}

\begin{figure}[ht]
    \centering
    \includegraphics[width=0.75\textwidth]{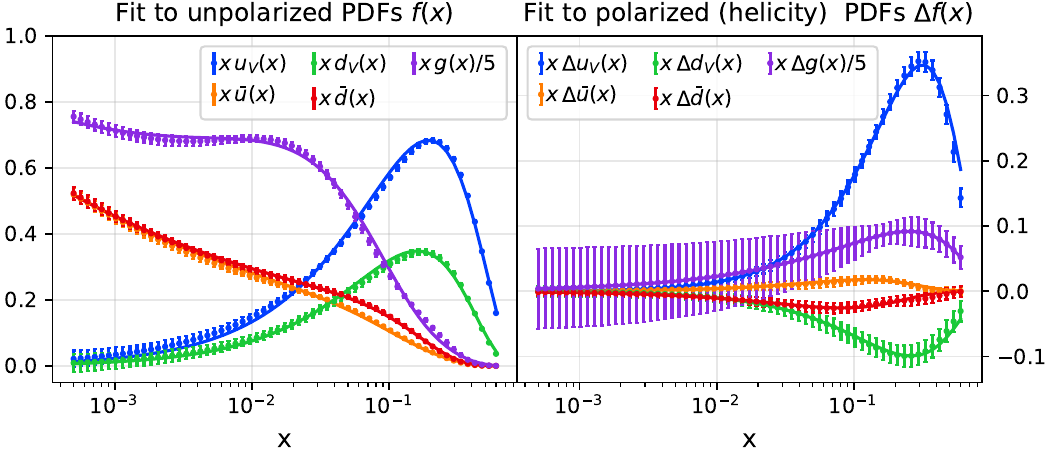}
    \includegraphics[width=0.75\textwidth]{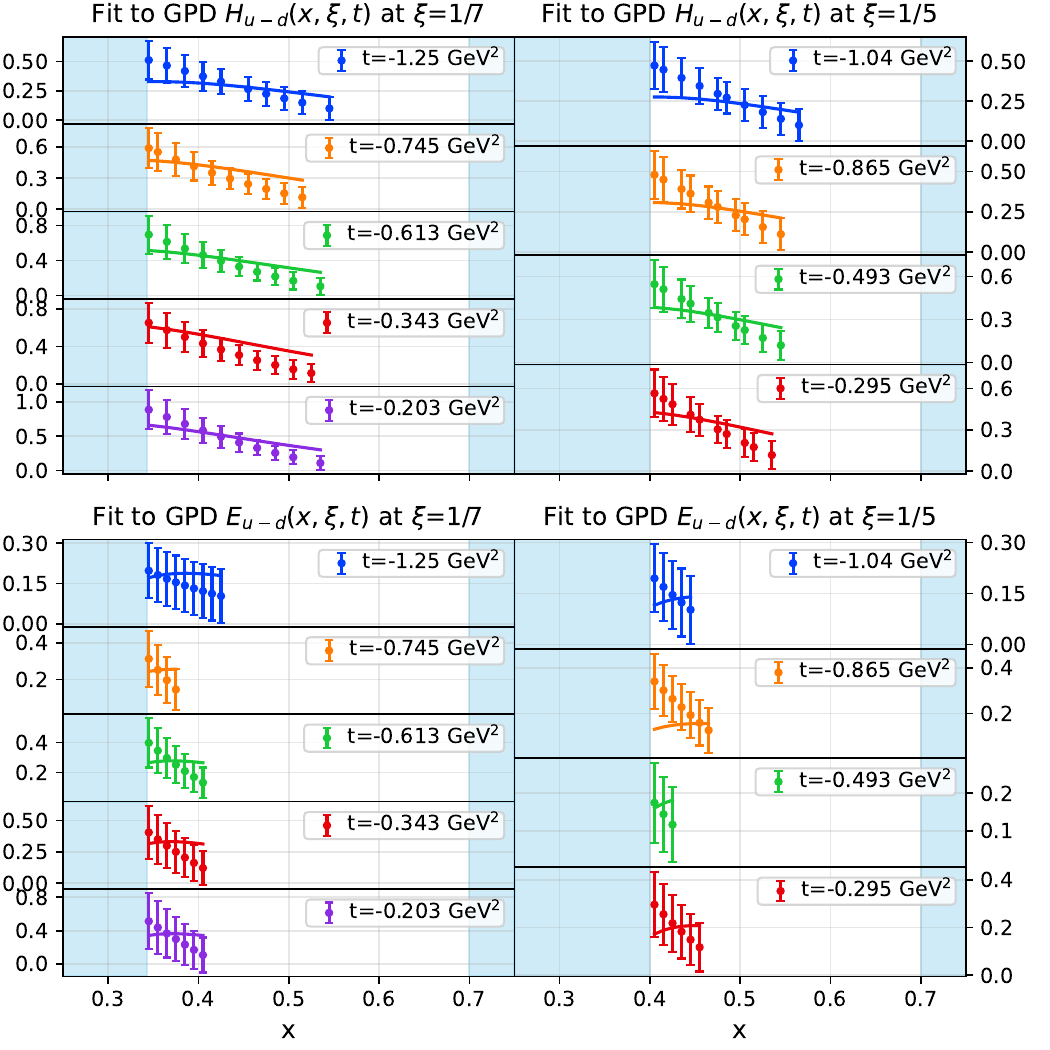}
    \caption{\raggedright Fits to the JAM polarized and unpolarized PDFs, as well as lattice-calculated GPDs at zero and nonzero skewness.}
    \label{fig:PDF}
\end{figure}

\begin{figure}[ht]
    \centering
    \includegraphics[width=0.75\textwidth]{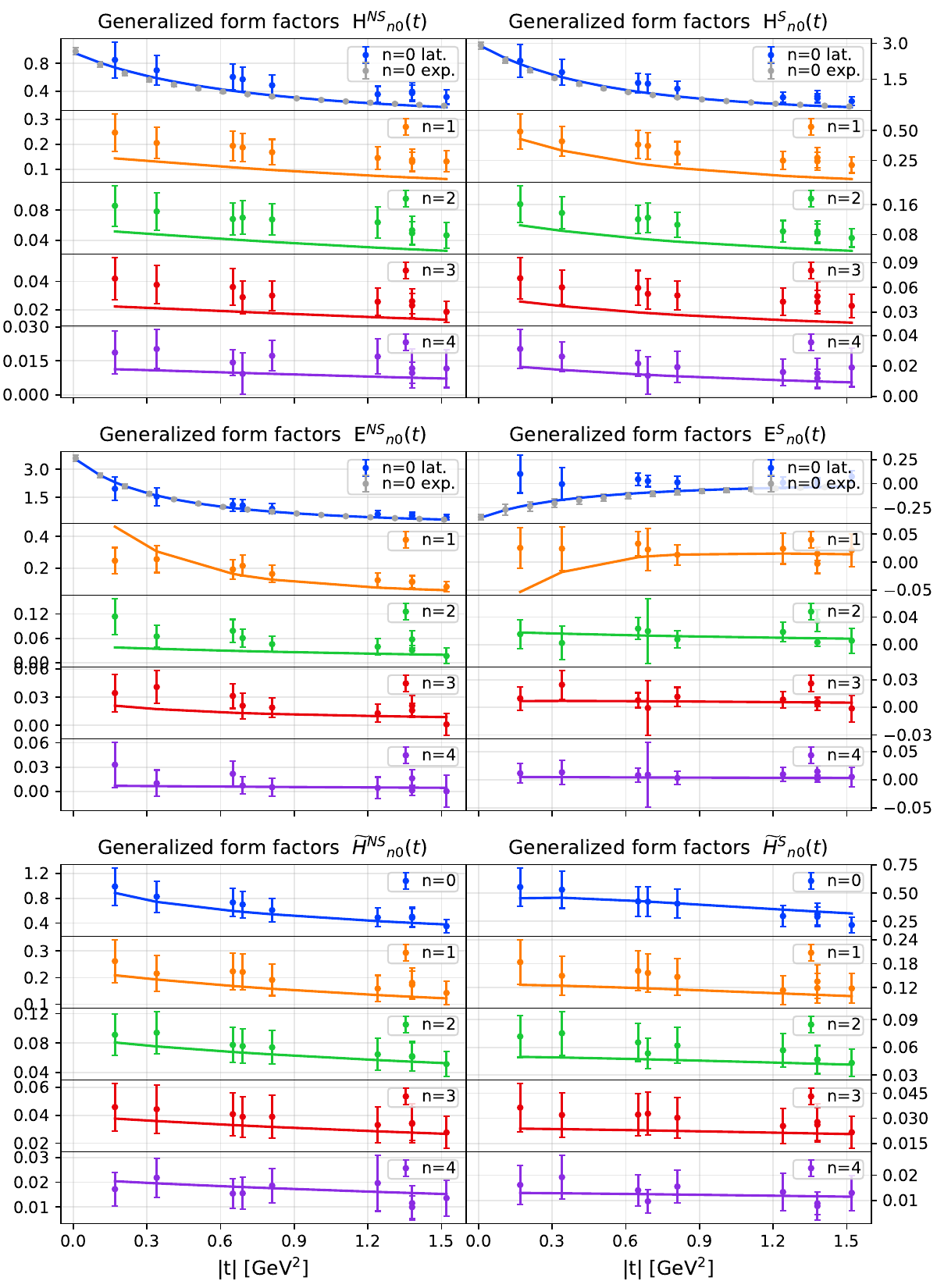}
    \includegraphics[width=0.85\textwidth]{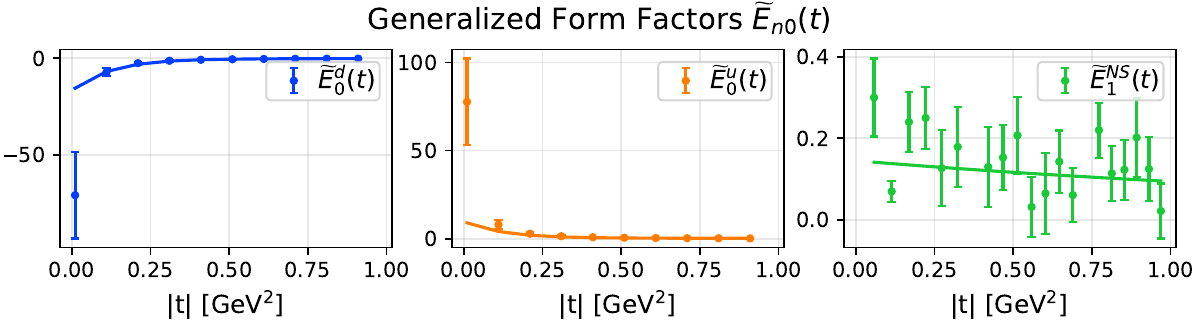}
    \caption{\raggedright Fits to the global charge form factors and lattice-calculated gravitational form factors.}
    \label{fig:GFF}
\end{figure}

\begin{figure}[ht]
    \centering
    \includegraphics[width=0.6\textwidth]{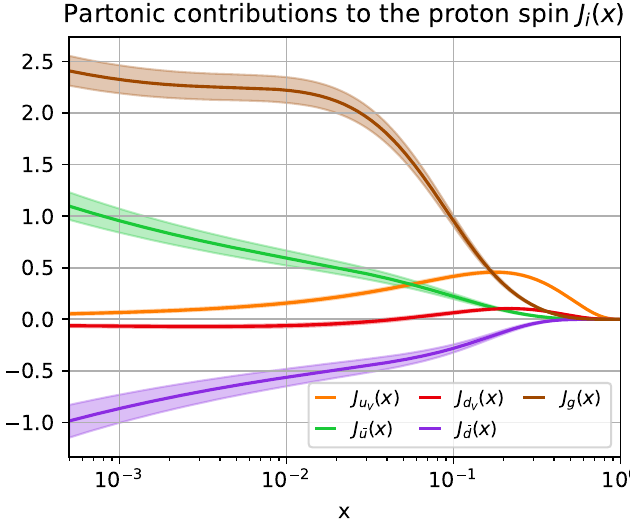}
    \caption{\raggedright Proton spin decomposition based on the GUMP1.0 GPDs at $\mu=2$ GeV. Bands correspond to 90\% confidence intervals considering only the Hessian errors.}
    \label{fig:spindecomp}
\end{figure}

Finally, the fit results are illustrated in Figs.~\ref{fig:DVCSUU}--\ref{fig:GFF}. We also present the partonic contributions to the proton in Fig.~\ref{fig:spindecomp}. The analysis provides a consistent description of DVCS cross sections and asymmetries measured at JLab, as well as DVCS and DV$\rho$P cross sections from HERA. It simultaneously accommodates constraints from global PDF fits and lattice QCD inputs on GPDs, both at zero and finite skewness. Finally, the framework reproduces nucleon form factors and lattice-calculated gravitational form factors, demonstrating the global consistency of the extraction.

We comment that most data are taken directly from the cited references, with a few modifications. First, all DV$\rho$P data are converted to longitudinal cross sections using the longitudinal-to-transverse ratio $R$ extracted from experimental measurements~\cite{ZEUS:2007iet, H1:2009cml}. Second, DVCS asymmetries measured at nearby kinematic points are clustered in $(x_B,t,Q)$ to avoid redundant calculations, with the cluster averages $(\bar{x}_B,\bar{t},\bar{Q})$ shown in the plots. Finally, lattice-calculated GPDs appear to turn negative and oscillate at large $x$, likely due to systematic uncertainties and the breakdown of large-momentum effective theory at modest proton momentum $P_z$. To mitigate this, we truncate inputs where $F(x,\xi,t)<f_0$ and set uncertainties as $\delta F'=\max(\delta F,f_0)$, with $f_0$ a cutoff parameter. A more detailed treatment is left for future work.

In Table. \ref{tab:besffit}, we also present all the best-fit parameters. During the analysis, we fixed some parameters, such as the Regge trajectory of the sea quarks and gluons $\alpha'$, as they are not as constrained by the observables in this work. We also reduce the number of $R_{\xi^4}$ coefficients for the axial-vector GPDs due to the lack of constraints. We use $b_{H_{\rm{sea}}}$ for the $b$ slope of up and down sea quarks and $b_{H_g}$ for gluons. For the axial-vector ones, we use $b_{\widetilde{H}_{\rm{sea}}}$ for all the quarks and gluons.

\begin{table}[ht]
    \centering
    \caption{\label{tab:besffit} A summary of the obtained best-fit parameters. }
    \renewcommand{\arraystretch}{1.5}
    \begin{tabular}
    {>{\centering\arraybackslash} p{0.09\textwidth} >{\centering\arraybackslash} p{0.14\textwidth} >{\centering\arraybackslash} p{0.09\textwidth} >{\centering\arraybackslash} p{0.14\textwidth} >{\centering\arraybackslash} p{0.09\textwidth} >{\centering\arraybackslash} p{0.14\textwidth} >{\centering\arraybackslash} p{0.09\textwidth} >{\centering\arraybackslash} p{0.14\textwidth}} 
    \multicolumn{8}{c}{Parameters and best-fit values for vector GPDs $H$ and $E$} \\ \hline \hline
    Name  & Value (error)  & Name  & Value (error)  &Name  & Value (error)  & Name  & Value (error) \\ \hline\hline
    $N^{H}_{u_V}$        & 0.3023(7)  & $\alpha^{H}_{u_V}$    & 0.227(5)   & $\beta^{H}_{u_V}$   & 3.219(18) & $\alpha'^{H}_{u_V}$  & 0.753(7) \\
    \hline
    $M^{-2}{}^{H}_{u_V}$ & 0.0000(15) & $N^{H}_{\bar u}$      & 0.0279(3)& $\alpha^{H}_{\bar u}$ & 1.200(4) & $\beta^{H}_{\bar u}$ & 6.10(11) \\
    \hline
    $\alpha'^{H}_{\bar q}$ & 0.15 (fixed) & $N^{H}_{\bar u,2}$ & 0.0022(2)& $\alpha^{H}_{\bar u,2}$ & 0.25(11) & $\beta^{H}_{\bar u,2}$ & 20.0(27) \\
    \hline
    $N^{H}_{d_V}$        & 0.1324(9)  & $\alpha^{H}_{d_V}$    & 0.201(12)  & $\beta^{H}_{d_V}$   & 4.20(6)  & $\alpha'^{H}_{d_V}$  & 0.47(5) \\
    \hline
    $M^{-2}{}^{H}_{d_V}$ & 0.46(8)    & $N^{H}_{\bar d}$      & 0.0210(6)  & $\alpha^{H}_{\bar d}$ & 1.188(7) & $\beta^{H}_{\bar d}$ & 9.7(4) \\
    \hline
    $N^{H}_{\bar d,2}$   & 0.0170(8)  & $\alpha^{H}_{\bar d,2}$ & 0.13(8)  & $\beta^{H}_{\bar d,2}$ & 9.9(5) & $N^{H}_{g}$           & 0.278(4) \\
    \hline
    $\alpha^{H}_{g}$     & 1.114(4)   & $\beta^{H}_{g}$       & 5.8(1)    & $\alpha'^{H}_{g}$    & 0.15(fixed) & $M^{-2}{}^{H}_{g}$ & 5.00(13) \\
    \hline
    $N^{H}_{g,2}$        & 0.1193(33) & $\alpha^{H}_{g,2}$    & 0.566(14) & $\beta^{H}_{g,2}$    & 20.0(3) & $N^{E}_{u_V}$         & 0.111(5) \\
    \hline
    $\alpha^{E}_{u_V}$   & 0.766(24)  & $\beta^{E}_{u_V}$     & 2.23(18)  & $\alpha'^{E}_{u_V}$  & 0.552(22) & $N^{E}_{d_V}$         & -0.059(4) \\
    \hline
    $\alpha^{E}_{d_V}$   & 0.846(13)  & $\beta^{E}_{d_V}$     & 4.09(29)  & $\alpha'^{E}_{d_V}$  & 0.363(14) & $R^{E}_{\bar u}$      & 3.22(30) \\
    \hline
    $R^{E}_{\bar d}$     & $-$4.80(28)  & $R^{E}_{g}$           & 0.30(4)   & $R^{H}_{u_V,{\xi^2}}$   & $-$1.577(27) & $R^{H}_{d_V,{\xi^2}}$    & $-$1.75(13) \\
    \hline
    $R^{H}_{g,{\xi^2}}$     & 0.06(6)    & $R^{E}_{u_V,{\xi^2}}$    & 0.72(8)   & $R^{E}_{d_V,{\xi^2}}$   & $-$10.0(3) & $R^{E}_{g,{\xi^2}}$      & 10.0(7) \\
    \hline
    $R^{H}_{u_V,{\xi^4}}$   & 0.336(7)   & $R^{H}_{d_V,{\xi^4}}$    & 0.245(35) & $R^{H}_{g,{\xi^4}}$     & 0.231(22) & $R^{E}_{u_V,{\xi^4}}$    & $-$0.066(23) \\
    \hline
    $R^{E}_{d_V,{\xi^4}}$   & 2.88(4)    & $R^{E}_{g,{\xi^4}}$      & $-$2.46(21) & $b_{H_\text{sea}}$    & 2.96(9)  & $b_{H_g}$       & 2.64(13) \\
    \hline
    \end{tabular}
    \begin{tabular}
    {>{\centering\arraybackslash} p{0.09\textwidth} >{\centering\arraybackslash} p{0.14\textwidth} >{\centering\arraybackslash} p{0.09\textwidth} >{\centering\arraybackslash} p{0.14\textwidth} >{\centering\arraybackslash} p{0.09\textwidth} >{\centering\arraybackslash} p{0.14\textwidth} >{\centering\arraybackslash} p{0.09\textwidth} >{\centering\arraybackslash} p{0.14\textwidth}} 
     \multicolumn{8}{c}{Parameters and best-fit values for axial vector GPDs $\widetilde{H}$ and $\widetilde{E}$} \\ \hline \hline
    Name  & Value (error)  & Name  & Value (error)  &Name  & Value (error)  & Name  & Value (error) \\ \hline\hline
    $N^{\widetilde H}_{u_V}$        & 0.1777(24) & $\alpha^{\widetilde H}_{u_V}$   & $-$0.209(21) & $\beta^{\widetilde H}_{u_V}$   & 2.60(7)   & $\alpha'^{\widetilde H}_{u_V}$ & 0.301(23) \\
    \hline
    $N^{\widetilde H}_{\bar u}$     & 0.0049(40) & $\alpha^{\widetilde H}_{\bar u}$ & 0.07(8)    & $\beta^{\widetilde H}_{\bar u}$ & 6.9(7)    & $\alpha'^{\widetilde H}_{\bar q}$ & 0.15(fixed) \\
    \hline
    $N^{\widetilde H}_{d_V}$        & $-$0.0499(26) & $\alpha^{\widetilde H}_{d_V}$  & 0.07(6)    & $\beta^{\widetilde H}_{d_V}$   & 2.62(27)  & $\alpha'^{\widetilde H}_{d_V}$  & 0.96(21) \\
    \hline
    $N^{\widetilde H}_{\bar d}$     & $-$0.0061(70) & $\alpha^{\widetilde H}_{\bar d}$ & 0.30(7)    & $\beta^{\widetilde H}_{\bar d}$ & 7.8(12)  & $N^{\widetilde H}_{g}$          & 0.053(5) \\
    \hline
    $\alpha^{\widetilde H}_{g}$     & 0.42(15)    & $\beta^{\widetilde H}_{g}$     & 1.7(5)     & $\alpha'^{\widetilde H}_{g}$  & 0.15(fixed) & $N^{\widetilde E}_{u_V}$        & 0.081(13) \\
    \hline
    $\alpha^{\widetilde E}_{u_V}$   & $-$2.0(5)     & $\beta^{\widetilde E}_{u_V}$   & 9.5(1.3)   & $\alpha'^{\widetilde E}_{u_V}$ & 0.32(16) & $N^{\widetilde E}_{d_V}$       & $-$0.064(13) \\
    \hline
    $R^{\widetilde E}_{\text{Sea}}$ & 100(6)      & $R^{\widetilde H}_{u_V,{\xi^2}}$   & $-$1.25(7)   & $R^{\widetilde H}_{d_V,{\xi^2}}$ & $-$4.1(8)   & $R^{\widetilde H}_{g,{\xi^2}}$    & 0(fixed) \\
    \hline
    $R^{\widetilde E}_{u_V,{\xi^2}}$   & $-$6.0(6)     & $R^{\widetilde E}_{d_V,{\xi^2}}$   & $-$5.2(16)   & $R^{\widetilde E}_{g,{\xi^2}}$   & 0(fixed) & $R^{\widetilde H}_{u_V,{\xi^4}}$ & 0(fixed) \\
    \hline
    $R^{\widetilde H}_{d_V,{\xi^4}}$   & 0(fixed) & $R^{\widetilde H}_{g,{\xi^4}}$    & 0(fixed) & $R^{\widetilde E}_{u_V,{\xi^4}}$ & 0(fixed) & $R^{\widetilde E}_{d_V,{\xi^4}}$ & 0(fixed) \\
    \hline
    $R^{\widetilde E}_{g,{\xi^4}}$     & 0(fixed) & $b_{\widetilde H _\text{sea}}$  & 7.9(5)     &  ~ & ~ & ~ & ~ \\
    \hline
    \end{tabular}
\end{table}
\end{document}